\documentclass[aps,reprint,prb,twocolumn]{revtex4-1}
\pdfoutput=1

\usepackage{graphicx}
\usepackage{amssymb}
\usepackage{amsmath}
\usepackage{amsthm}
\usepackage{slashed}
\usepackage[colorlinks=true, linkcolor=blue, citecolor=blue, urlcolor=blue]{hyperref}
\usepackage{color,rotating}
\usepackage{bbm}
\usepackage{array}
\usepackage{pstricks}
\usepackage{pstricks-add}
\usepackage{colonequals}
\usepackage[utf8]{inputenc}

\DeclareMathAlphabet{\EuRoman}{U}{eur}{m}{n}
\SetMathAlphabet{\EuRoman}{bold}{U}{eur}{b}{n}

 \graphicspath{{./figures_HB/}}

\newcommand{\UV}{{\small UV}}
\newcommand{\IR}{{\small IR}}
\newcommand{\FRG}{{\small FRG}}
\newcommand{\RG}{{\small RG}}

\newcommand{\NLO}{{\small NLO}}

\newcommand{\PMS}{{\small PMS}}
\newcommand{\MC}{{\small MC}}
\newcommand{\CBS}{{\small CBS}}

\newcommand{\eg}{{\textit{e.g.}}}
\newcommand{\ie}{{\textit{i.e.}}}

\allowdisplaybreaks

\begin{document}

\title{Critical (Chiral) Heisenberg Model with the Functional Renormalisation Group}

\author{Benjamin Knorr}
\email[Electronic address: ]{b.knorr@science.ru.nl}
\affiliation{Theoretisch-Physikalisches Institut, Universit\"at Jena,
Max-Wien-Platz 1, 07743 Jena, Germany}
\affiliation{Institute for Mathematics, Astrophysics and Particle Physics (IMAPP), \\
Radboud University Nijmegen, Heyendaalseweg 135, 6525 AJ Nijmegen, The Netherlands}

\begin{abstract}
We discuss the Heisenberg model and its chiral extension in an extended truncation with the help of functional methods.
Employing computer algebra to derive the beta functions, and pseudo-spectral methods to solve them, we are able
to go significantly beyond earlier approximations, and provide new estimates on the critical quantities of both models.
The fixed point of the Heisenberg model is mostly understood, and our results are in agreement with estimates from various
other approaches, including Monte Carlo and conformal bootstrap studies. By contrast, in the chiral case, the formerly known disagreement with lattice studies 
persists, raising the question whether actually the same universality class is described.
\end{abstract}

\maketitle

\section{Introduction}

Many magnetic materials can be efficiently described by the Heisenberg model, which consists
of a vector invariant under $O(3)$ rotations. Examples are the Curie transition in isotropic
ferromagnets and antiferromagnets at the N\'eel transition point \cite{Pelissetto:2000ek}. The price to pay for the simplicity
of this model is the negligence of some interactions that are present in real materials, for example
dipolar interactions. Even though such interactions can be relevant perturbations \cite{PhysRevB.8.3323,
Carmona:1999rm, Fisher:1974uq}, studies show that
their impact is small \cite{Aharony:1974zz, Calabrese:2002sz, Carmona:1999rm, PhysRevB.62.11649}.

Another interesting case where the Heisenberg model, and its chiral extension,
play a role is the description of graphene \cite{Novoselov2005, Geim2007, CastroNeto:2009zz, RevModPhys.83.407},
and related materials \cite{PhysRevB.76.075131, APL:1.3419932,
PhysRevLett.108.155501, doi:10.1021/nn504451t, PhysRevLett.102.236804, doi:10.1021/nn4009406, doi:10.1021/acs.nanolett.5b00085,
PhysRevLett.101.126804, Guinea5391, PhysRevB.86.081405, Yankowitz2012, Ponomarenko2013, Hunt1427,
SoltanPanahi2011, PhysRevLett.110.106801, PhysRevB.93.134502, PhysRevB.79.241406, Tarruell2012, Polini2013, PhysRevLett.108.140405,
Liu864, Neupane2014, PhysRevLett.113.027603}. 
Graphene is a very interesting material. Due to its honeycomb structure, it behaves very differently when compared with
standard materials with a Bravais lattice structure.
A direct consequence of the lattice structure of graphene is that the Fermi surface consists of two points only,
and in principle invalidates the Landau Fermi liquid construction. 
Expanding the dispersion relation around the Fermi points, continuum models for the fermionic non-interacting low-energy excitations
with relativistic symmetry can be constructed \cite{doi:10.1080/00018732.2014.927109, PhysRevB.80.081405,
PhysRevB.86.121402, PhysRevB.66.045108, PhysRevB.75.134512, PhysRevLett.97.146401, PhysRevLett.100.146404, Herbut:2009vu,
PhysRevLett.100.156401, PhysRevB.87.085136, PhysRevB.89.035103, PhysRevB.79.085116, Janssen:2014gea, PhysRevLett.98.186809,
PhysRevB.82.035429, PhysRevB.85.155439, PhysRevB.90.035122, PhysRevB.92.155137, PhysRevLett.106.236805, RevModPhys.84.299,
PhysRevLett.86.958, Gehring:2015vja, PhysRevB.78.165423, PhysRevB.81.125105, 1751-8121-45-38-383001, PhysRevB.81.085105, PhysRevB.87.041401,
PhysRevB.82.205106, PhysRevD.88.065008, Jakovac2015, Scherer:2016zwz, Choi:2016sxt, Mihaila:2017ble, Classen:2017hwp, Iliesiu:2017nrv}.

A particular model for Dirac materials is given in Refs. \onlinecite{Roy:2011pg, Roy:2013aya, Classen:2015mar}. In Ref. \onlinecite{Knorr:2016sfs}, we dealt 
with the Ising-like subset of this model, which corresponds to a 3d Gross-Neveu model for four-component Dirac fermions in a reducible representation. 
Supersymmetric aspects of this model are considered \eg{} in Refs. \onlinecite{Heilmann:2014iga, Hellwig:2015woa, Gies:2017tod}.
The aim of the present work is to consider the Heisenberg-like subset, where an $O(3)$-invariant vector is coupled to these fermions via Pauli matrices.

Monolayered graphene is an effectively (2+1)-dimensional material. Since the upper (lower) critical dimension of
the model that we consider here is $d=4 (2)$, the accuracy of perturbative results obtained with $\epsilon$-expansions \cite{Rosenstein:1993zf, Janssen:2014gea}
around one of the critical dimensions is an open question. On the other hand, lattice studies involving fermions might
suffer from sign problems. Here, we will treat our model with the continuous realisation of the exact renormalisation group
by Wetterich \cite{Wetterich:1992yh}.

There are several difficulties one encounters in the study of critical phenomena with functional methods.
The first step is to decide on an approximation (often called truncation), and to determine
the renormalisation group (\RG{}) flow of the operators present in this truncation. We will do this in complete analogy to
the earlier study of the Ising counterpart \cite{Knorr:2016sfs} of this work with the help of xAct \cite{xActwebpage, 2007CoPhC.177..640M,
2008CoPhC.179..586M,2008CoPhC.179..597M,2014CoPhC.185.1719N}. Consequently, the resulting differential
equations have to be solved. We will use pseudo-spectral methods to do so, which were systematically put forward
in the present context in Refs. \onlinecite{Borchardt:2015rxa, Borchardt:2016pif},
and applications of these methods to functional renormalisation group (\FRG{}) studies
can be found in Refs. \onlinecite{Heilmann:2014iga, Borchardt:2016xju, Borchardt:2016kco, Knorr:2016sfs}.

The aim of the present work is to provide new estimates on critical quantities for both the Heisenberg model
and its chiral equivalent. Several methods agree quite well on the Heisenberg model,
whereas for the model involving fermions, the situation is not settled. In particular,
there is a clash between \FRG{} results and quantum Monte Carlo simulations on the values of
the relevant critical exponent and the bosonic anomalous dimension \cite{Janssen:2014gea}. Our study shows that this clash
persists, even when a truncation retaining 12 operators is considered.

This paper is organised as follows. We start with a short recap on the \FRG{} in \autoref{sec:FRG},
followed by the introduction of our model in \autoref{sec:model}. Subsequently, we discuss the results
of the model, first without fermions in \autoref{sec:HB}, then with fermions in \autoref{sec:GN}.
We finally summarise the results in \autoref{sec:summary}.

For the reader less interested in the technical details, in \autoref{tab:summ_res} we summarise the numerical findings and compare to literature values for both the Heisenberg and the chiral Heisenberg model.

\begin{table*}
\begin{tabular}{cccc}
\hline\hline
  Heisenberg model & $\theta_1$ & $\eta $ & \\ \hline
  FRG (this work) & 1.4178 & 0.04662 &  \\
  FRG \cite{Classen:2015mar} & 1.359 & 0.041 & \\
  Monte Carlo \cite{PhysRevB.84.125136} & 1.4053(20) & 0.0378(3) & \\
  conformal bootstrap \cite{Kos2016} & 1.4043(55) & 0.03856(124) & \\ \hline\hline
  chiral Heisenberg model& $\theta_1$ & $\eta_\phi$ & $\eta_\psi$ \\ \hline
  FRG (this work) & 0.795 & 1.032 & 0.071(2) \\
  FRG \cite{Janssen:2014gea} & 0.772 & 1.015 & 0.084 \\
  quantum Monte Carlo \cite{Toldin:2014sxa} & 1.19(6) & 0.70(15) & \textemdash \\
  quantum Monte Carlo \cite{Otsuka:2015iba} & 0.98(1) & 0.49(5) & 0.20(2) \\
  $(4-\epsilon)$ 2nd order \cite{Rosenstein:1993zf, Janssen:2014gea} & 0.834 & 0.959 & 0.242 \\
  $(4-\epsilon)$ 4th order (3/1) Pad\'e \cite{Zerf:2017zqi} & 0.645 & 0.956 & 0.156 \\ \hline\hline
\end{tabular}
\caption{Comparison of the results on the first critical exponent and the anomalous dimension(s) of the Heisenberg and chiral Heisenberg model with the literature.
The results for the Heisenberg model are in reasonable agreement with Monte Carlo and conformal bootstrap techniques. In the chiral case, the situation
is not yet settled, and different methods disagree by factors of up to three on different quantities. To estimate the bosonic 
anomalous dimension from the results of Ref. \onlinecite{Otsuka:2015iba}, we employed the hyperscaling relation $\eta_\phi = 2\beta\theta_1+2-d$.}
\label{tab:summ_res}
\end{table*}

\section{Functional renormalisation group}\label{sec:FRG}

A convenient way to investigate quantum fluctuations in a non-perturbative manner is
the effective average action, $\Gamma_k$. It interpolates between the classical action
and the full quantum effective action, and fulfils the functional equation \cite{Wetterich:1992yh}
\begin{equation} \label{eq:Wetterich}
 \partial_t \Gamma_k = \frac{\mathbf{i}}{2} \mathrm{STr}\left[ \left(\Gamma^{(2)}_k + R_k \right)^{-1} (\partial_t R_k) \right], \quad 
t=\ln\bigg(\frac{k}{\Lambda}\bigg),
\end{equation}
where $\Gamma^{(2)}_k$ is the second functional derivative of the effective average action
with respect to the fields that are considered, and $t$ is the \RG{} ``time'', measuring momenta
in units of a fixed momentum scale $\Lambda$. The super-trace STr indicates a sum over discrete
and an integration over continuous indices, and includes a minus sign for fermions. The equation is well-defined
due to the regulator $R_k$, which provides both ultraviolet (\UV{}) and infrared (\IR{}) regularisation.
For reviews on the \FRG{}, see Refs. \onlinecite{Berges:2000ew,Pawlowski:2005xe,Gies:2006wv,Kopietz:2010zz,Delamotte:2007pf}.

Solving the flow equation \eqref{eq:Wetterich} exactly is typically very difficult, and approximations have
to be introduced. As an example, in gauge theories it is often important to resolve the full
momentum dependence of vertices, thus a vertex expansion is employed. By contrast, in scalar and fermionic models,
it seems to be the case that retaining momenta up to a fixed power, but arbitrary field dependence, captures
the most important physics. In the present work, we will investigate an approximation which retains
operators with at most two derivatives and two fermion fields. In the bosonic case, all operators of this class are included, and this approximation is 
commonly called next-to-leading order (\NLO{}) in a derivative expansion. By contrast, in the chiral case, the full \NLO{} approximation is intricate due to 
the complicated tensor structure, thus we will focus on a subset of all possible operators, concentrating on momentum-dependent corrections to the Yukawa 
coupling. We further discuss the effect of two rank 2 tensor couplings, showing that their impact is suppressed.

\section{The model}\label{sec:model}

The model that we will describe shares essential features with classical four-fermion models such as the Gross-Neveu model \cite{Gross:1974jv}.
We will deal with two flavours of massless relativistic Dirac fermions in a four-dimensional reducible
representation. In the conventions of Ref. \onlinecite{Knorr:2016sfs}, the Minkowskian microscopic action
of this model reads
\begin{equation}
 S = \int \left( \overline\psi \left( \mathbbm 1_2 \otimes \slashed{\partial} \right) \psi 
 + \frac{\overline g}{4} \left( \overline\psi \left( \sigma^a \otimes \mathbbm 1_4 \right) \psi \right)^2 \right) \, ,
\end{equation}
where $\sigma^a$ are the Pauli matrices. The crucial symmetry of this theory is the invariance under $SU(2)$ spin rotations. By a partial bosonisation, we can 
reformulate the action in terms of a Yukawa theory with action
\begin{equation}
 S^\text{pb} = \int \left( \overline \psi \left( \left( \mathbbm 1_2 \otimes \slashed{\partial} \right) 
 + \overline{h} \, \phi^a \left( \sigma^a \otimes \mathbbm 1_4 \right) \right) \psi - \overline m^2 (\phi^a)^2 \right) \, .
\end{equation}
Here, $\overline g = \overline h^2/\overline m^2$, and $\phi^a$ is a vector field invariant under $SU(2) \simeq O(3)$ rotations.
This will be the starting point for our investigation. In general, once quantum fluctuations are included,
all further operators that are allowed by the symmetries will be generated, and have to be taken into account. In the following,
we will include operators with at most two fermions and two derivatives.

Let us start with the purely bosonic part of our ansatz for the effective average action,
\begin{equation}
 \Gamma_k^\text{bos} = \int \left( \frac{1}{2} Z_\phi(\rho) \left(\partial_\mu \phi^a\right)^2
 + \frac{1}{2} Y_\phi(\rho) \left(\partial_\mu \rho\right)^2 - V(\rho) \right) \, ,
\end{equation}
which includes the wave function renormalisation $Z_\phi$, a correction term to the radial propagator, $Y_\phi$,
and the potential $V$. We also introduced $\rho = \phi^a \phi^a/2$ for convenience. In the subsequent section,
where we discuss the Heisenberg model, this ansatz will be discussed.

For the fermions, we first introduce the kinetic term with fermion wave function renormalisation $Z_\psi$, and the standard Yukawa coupling, $g_1$,
\begin{widetext}
\begin{equation}\label{eq:FermKin}
 \Gamma_k^\text{ferm} = \int \left( \frac{1}{2} Z_\psi(\rho) \left(  \overline{\psi} \left( \mathbbm 1_2 \otimes \slashed{\partial} \right) \psi 
 - (\partial^\mu \overline{\psi}) \left( \mathbbm 1_2 \otimes \gamma_\mu \right) \psi \right) + g_1(\rho) \phi^a \overline{\psi} \left( \sigma^a \otimes 
\mathbbm 1_4 \right) \psi \right) \, .
\end{equation}
There are 7 further operators that we will consider here. Most come with the tensor structure $\sigma^a \otimes \mathbbm 1_4$ and carry two derivatives, 
thus they are momentum-dependent extensions of the Yukawa coupling and will be labelled by a $g$. To study the effect of tensorial interactions, we further 
study two operators which couple via $\Sigma_{\mu\nu} = 2 [\gamma_\mu, \gamma_\nu]$:
\begin{equation}\label{eq:HGNint}
\begin{aligned}
 \Gamma_k^\text{int} &= \int \left[
 -\left( g_2(\rho) - \frac{1}{2} g_6(\rho) \right) \left( \partial^2 \phi^a \right) \overline \psi \left( \sigma^a \otimes \mathbbm 1_4 \right) \psi
 -\left( g_3(\rho) - \frac{1}{2} g_6'(\rho) \right) \left(\partial^2 \rho\right) \phi^a \overline \psi \left( \sigma^a \otimes \mathbbm 1_4 \right) \psi 
\right. \\
 &\qquad\quad +\frac{1}{2} g_4(\rho) \left(\partial^\mu \phi^b\right)^2 \phi^a \overline \psi \left( \sigma^a \otimes \mathbbm 1_4 \right) \psi
 +\left\{ g_5(\rho) - g_2'(\rho) - g_3(\rho) + g_6'(\rho) \right\} \left(\partial_\mu \phi^a\right)\left(\partial^\mu \rho\right) \overline \psi \left( 
\sigma^a \otimes \mathbbm 1_4 \right) \psi \\
 &\qquad\quad -\frac{1}{2} g_6(\rho) \phi^a \left( \overline \psi \left( \sigma^a \otimes \mathbbm 1_4 \right) \partial^2 \psi + \left( \partial^2 
\overline \psi \right) \left( \sigma^a \otimes \mathbbm 1_4 \right) \psi \right) \\
 &\qquad\quad \left. +\frac{1}{2} \left\{ 
T_1(\rho) \left( \partial_\mu \phi^a \right) + T_2(\rho) \phi^a \left(\partial_\mu \rho\right)\right\} \left( \overline \psi \left( \sigma^a \otimes 
\Sigma^{\mu\nu} \right) \partial_\nu \psi - \left( \partial_\nu \overline \psi \right) \left( \sigma^a \otimes \Sigma^{\mu\nu} \right) \psi \right) \right] \, .
\end{aligned}
\end{equation}
\end{widetext}
The specific linear combinations in front of the invariants are chosen for convenience, and simplify the calculation.
Our conventions on the Clifford algebra are the same as in Ref. \onlinecite{Knorr:2016sfs}. The full ansatz for the chiral Heisenberg model combines
all of this,
\begin{equation}\label{eq:HGNansatz}
 \Gamma_k^\text{HGN} = \Gamma_k^\text{bos} + \Gamma_k^\text{ferm} + \Gamma_k^\text{int} \, .
\end{equation}
All functions depend on the renormalisation group scale $k$, and have to be real
in order that the Minkowskian ansatz for the action is real. All algebraic manipulations are done in Minkowski space,
and only the final integration over the loop momentum is done after a Wick rotation.
The symmetries of the above model are discussed in
Ref. \onlinecite{Janssen:2014gea}, and the most constraining symmetry for the construction of the ansatz is the discrete $\mathbbm Z_2$
reflection symmetry,
\begin{equation}
 \psi \to (\mathbbm{1}_{N_f} \otimes \gamma_2) \psi, \, \overline\psi \to -\overline\psi (\mathbbm{1}_{N_f} \otimes \gamma_2), \, \phi^a \to -\phi^a \, ,
\end{equation}
together with a parity transformation of spacetime, \eg{} $x_1 \to -x_1$.
We don't expect further accidental symmetries as in the case of the chiral Ising model \cite{Knorr:2016sfs}, where a symmetry related to a reality constraint 
is present, constraining the occurrence of a certain operator class. This expectation comes from the explicit occurrence of a factor of $\mathbf{i}$ in 
the commutator of the Pauli matrices.

The truncation \eqref{eq:HGNansatz} goes significantly beyond any \FRG{} calculation of this model. All calculations so far only included
a field-dependent potential together with field independent but scale-dependent wave functions $Z_\phi, Z_\psi$ and Yukawa coupling $g_\phi$ 
\cite{Janssen:2014gea,Classen:2015mar}.

To discuss the critical behaviour of a given model, dimensionless or renormalised quantities are introduced. Fixed points, which describe \eg{} phase 
transitions,
are then characterised by the vanishing of the flow of these renormalised couplings. The relation between bare and renormalised quantities is straightforward, 
and
we will not write it down explicitly. It is in complete analogy to Ref. \onlinecite{Knorr:2016sfs}, and we encounter the same ambiguity: at which 
$\rho = \overline\rho$ do we normalise
the wave function renormalisations? Possible choices include the vacuum expectation value (vev), or zero. This ambiguity can be used to check the stability of 
our results. The running
of this normalisation is encoded in the anomalous dimensions,
\begin{align}
 \eta_\phi &= -\partial_t \ln Z_\phi(\overline\rho) \, , \notag \\
 \eta_\psi &= -\partial_t \ln Z_\psi(\overline\rho) \, .
\end{align}

Now, let us specify the regulator that we employ. In complete analogy to Ref. \onlinecite{Knorr:2016sfs},
we regularise the action by adding
\begin{align}
 \Delta S_\chi &= \int \left( \frac{1}{2} \phi^a \, R_\phi\left(\frac{p^2}{k^2}\right)\, \phi^a \right. \notag \\
 &\left.\qquad\qquad+ \overline{\psi}\, R_\psi\left(\frac{p^2}{k^2}\right)\, \frac{\left(\mathbbm{1}_{2}\otimes\gamma_\mu\right)\partial^\mu}{p} \psi  \right) 
\, .
\end{align}
Here, momentum arguments are to be understood as those after Wick rotation. To be able to optimise results with the principle of minimum sensitivity (\PMS{}), 
we will employ several regulator kernels. Optimisation aspects in the context of the \FRG{} are discussed \eg{} in Refs. \onlinecite{Litim:2000ci, 
Litim:2001up, Litim:2001fd, Litim:2002cf, Canet:2002gs, Canet:2003qd, Canet:2004xe, Litim:2005us, Pawlowski:2005xe, Fischer:2008uz, Pawlowski:2015mlf}. On the 
one hand, we discuss the linear regulator \cite{Litim:2001up},
\begin{align}
 R_\phi(x) &= k^2 (1-x) \, \theta(1-x) \, , 
\end{align}
with $\theta$ being the Heaviside step function. On the other hand, we also study a one-parameter family of exponential regulators given by
\begin{align}
 R_\phi(x) &= \frac{k^2}{2e^{x^a} - 1} \, , \notag \\
 R_\psi(x) &= \frac{k}{2e^{x^a} - 1} \, .
\end{align}
The numerical integration of the threshold functions is performed via an adaptive Gauss-Kronrod 7-15 rule, with the same parameters as chosen in Ref. 
\onlinecite{Knorr:2016sfs}.

\begin{figure}[!ht]
 \includegraphics[width=\columnwidth]{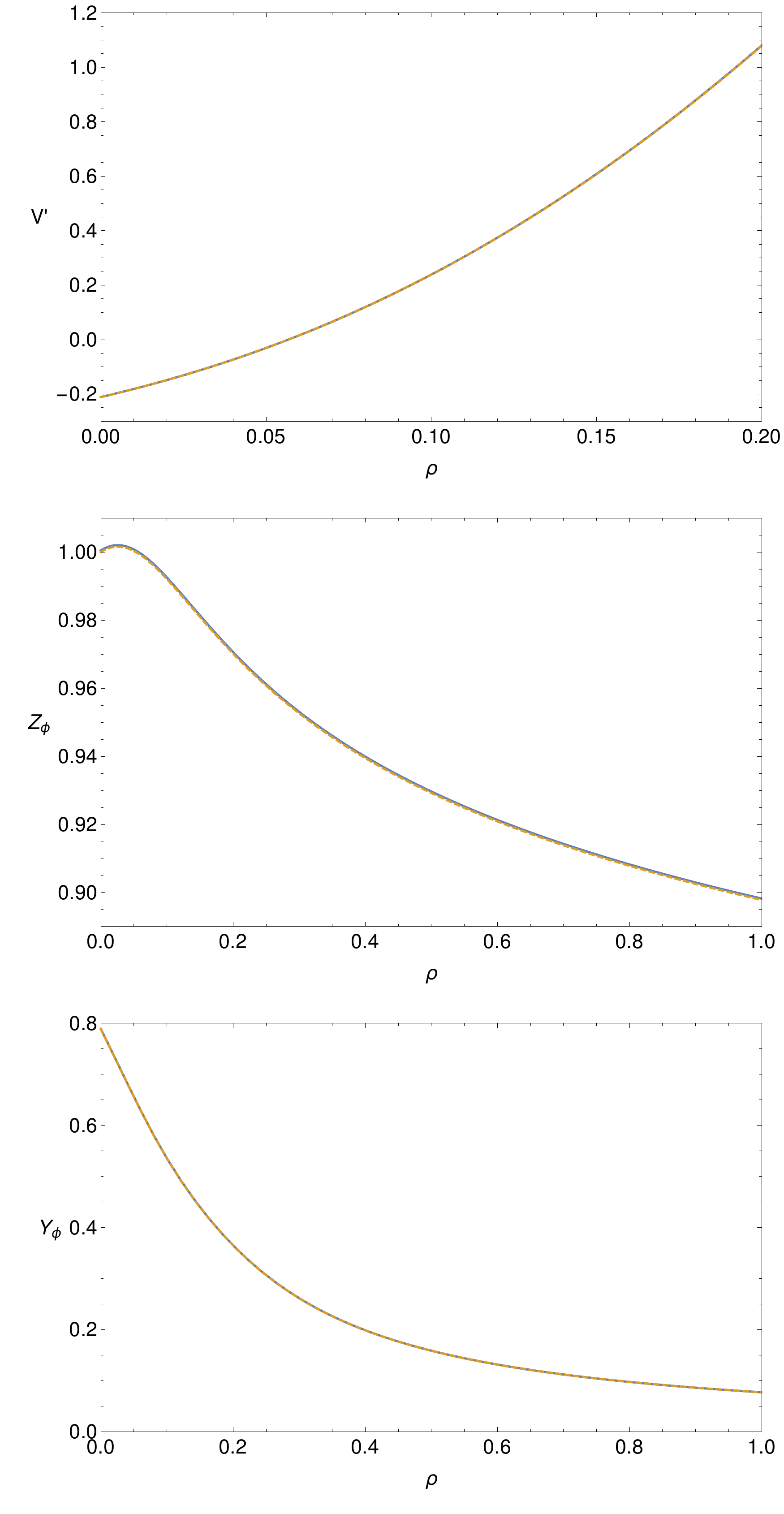}
 \caption{Fixed point solution to the Heisenberg model in \NLO{}. Shown are
 the derivative of the effective potential (top), the wave function renormalisation (middle),
 and the correction to the radial propagator (bottom). Blue solid lines correspond to
 the solution where $Z_\phi(\rho_0)=1$, whereas orange, dashed lines indicate that
 $Z_\phi(0)=1$ was chosen.
 In contrast to the Ising model, here both projection schemes deliver
 virtually the same fixed point solution. Even the small shift between the two
 wave function renormalisations can hardly be seen.}
 \label{fig:HB_NLO}
\end{figure}

Finally, some words on the derivation of the actual flow equations are in order. Clearly, it is very tedious to calculate the flow equations for all 
functions in \eqref{eq:HGNansatz} by hand. To minimise the danger of errors, we used the Mathematica package xAct \cite{xActwebpage, 2007CoPhC.177..640M,
  2008CoPhC.179..586M,2008CoPhC.179..597M,2014CoPhC.185.1719N} to derive them. The calculation proceeds similarly to the standard Gross-Neveu model 
\cite{Knorr:2016sfs},
except that the additional $SU(2)$ index makes the tensor structure richer, and thus the flavour structure cannot be treated abstractly. Still, in principle the 
calculation is
straightforward, but very lengthy \footnote{A notebook containing the explicit flow equations is available from the author upon request.}.

The system of 12 flow equations for the ansatz \eqref{eq:HGNansatz} has been solved with pseudo-spectral methods, which were systematically adapted to the 
present case in Ref. \onlinecite{Borchardt:2015rxa}, applications in the context of the \FRG{} can be found in Refs. \onlinecite{Heilmann:2014iga, 
Borchardt:2016pif, Borchardt:2016xju, Borchardt:2016kco, Knorr:2016sfs}. For the handling of linear algebra, we employed the library Eigen \cite{eigen}.

\section{Results for the Heisenberg model}\label{sec:HB}

We will now discuss the results for the Heisenberg model, \ie{} we switch off the fermions.
First the result obtained with the Litim regulator will be presented, and afterwards
optimised values with the help of the exponential regulators are given.

From here on, we discuss dimensionless quantities only, which are obtained by suitable rescalings with the \RG{} scale $k$. In the following, $\rho_0$ will 
denote the vev, so that $V'(\rho_0)=0$.

\begin{figure}
 \includegraphics[width=\columnwidth]{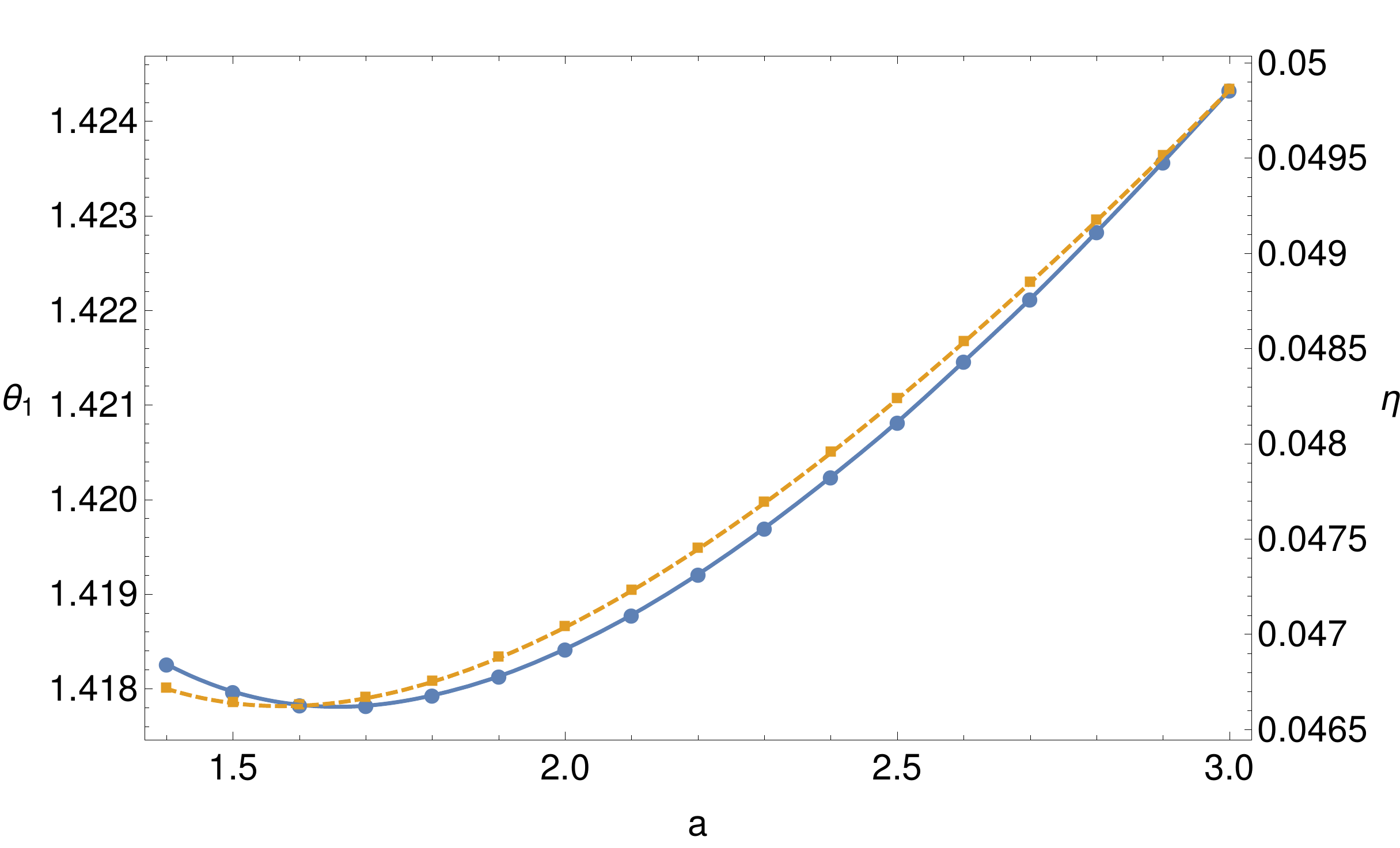}
 \caption{Dependence of the first critical exponent $\theta_1$ (blue dots) and the anomalous dimension
 $\eta$ (orange boxes) of the Heisenberg model on the regulator parameter $a$. An interpolation helps to guide the eye.}
 \label{fig:HB_NLO_opt}
\end{figure}

Two possible solutions are compared, where we use the aforementioned ambiguity in normalising the wave function
renormalisation to one at an arbitrary point. For scheme A, we fix $Z_\phi$ to unity at the vev, \ie{} $\bar \rho = \rho_0$,
whereas for scheme B, we fix $Z_\phi(0)=1$, such that $\bar\rho = 0$. The nomenclature follows Ref. \onlinecite{Knorr:2016sfs}.

All three fixed point functions are shown in \autoref{fig:HB_NLO}. In contrast to the case of the Ising model \cite{Knorr:2016sfs},
for the Heisenberg model, both projection schemes A and B deliver a consistent picture.
A small difference can only be seen (naturally) in the wave function renormalisation. This also settles in the
values for the vev and the anomalous dimension,
\begin{align}
 \rho_0^A &= 0.056838 \, , \qquad &\eta^A = 0.052347 \, , \notag \\
 \rho_0^B &= 0.056868 \, , \qquad &\eta^B = 0.052356 \, .
\end{align}
The difference between the two projections is at the per mille level.

Let us now discuss the critical exponents. In the context of the \FRG{}, they are determined as (minus) the eigenvalues
of the differential operator which is obtained by linearising the flow equations around the fixed point. Since trivial rescalings
of the field are of no physical interest, we further have to demand that the variation of the wave function renormalisation
at $\bar \rho$ vanishes.
The above picture carries over to the critical exponents, both schemes deliver quantitatively well agreeing values,
\begin{align}
 \theta_1^A &= 1.42965 \, , \qquad &\theta_1^B &= 1.42987 \, , \notag \\ 
 \theta_2^A &= -0.73398 \, , \qquad &\theta_2^B &= -0.73358 \, ,
\end{align}
the difference being in the sub per mille level.

\begin{figure*}
 \includegraphics[width=\textwidth]{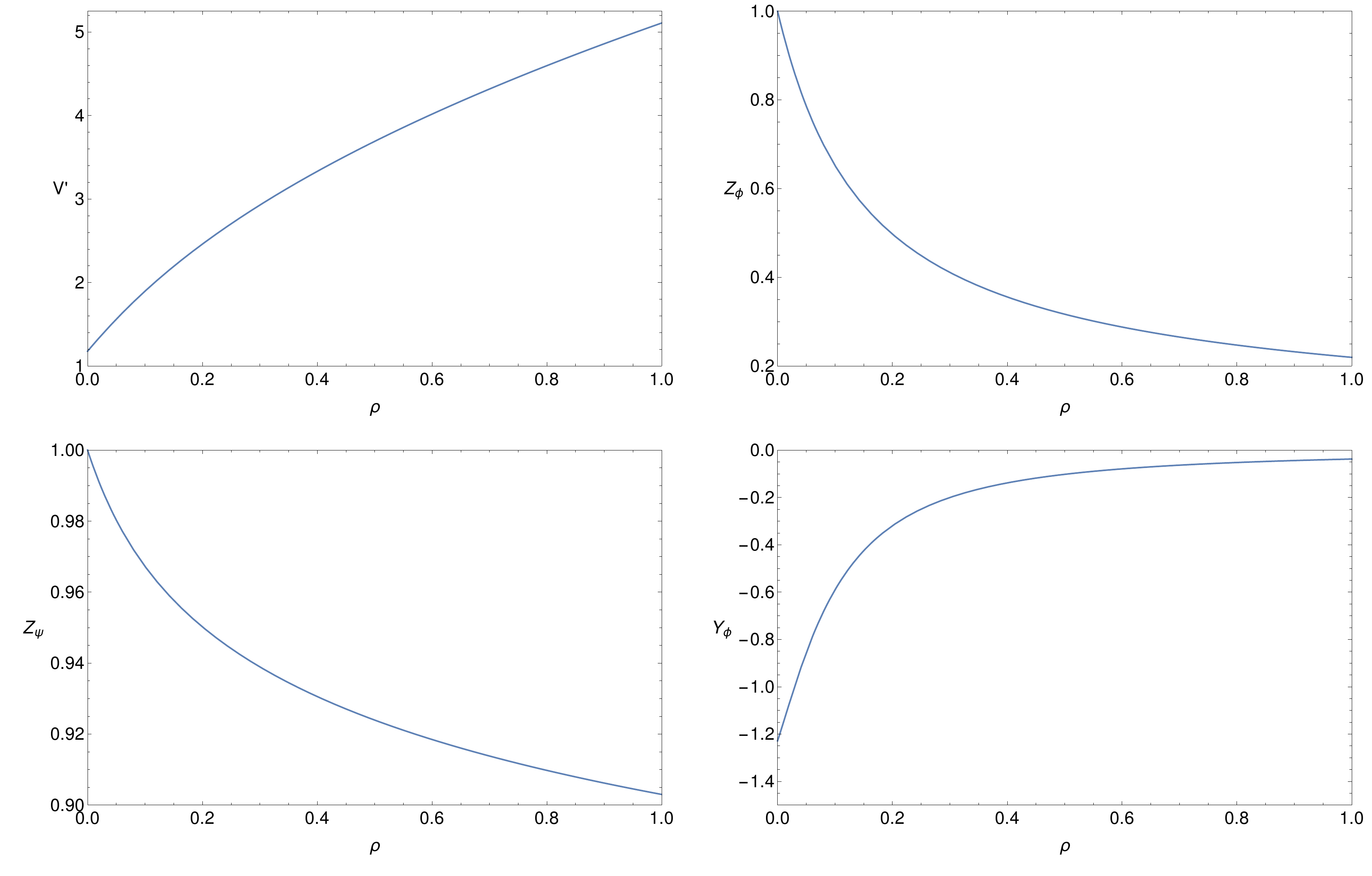}
 \caption{The first derivative of the potential and the wave function renormalisations
 at the fixed point of the chiral Heisenberg model, for the regulator parameter $a=2$. In contrast to the Heisenberg
 model, the correction to the radial wave function, $Y_\phi$, is negative. Still, the combination
 of $Z_\phi$ and $Y_\phi$ that appears in the propagator is strictly positive, and thus
 the propagator is well-defined for all field values.}
 \label{fig:cHB1}
\end{figure*}

Let us switch to the one-parameter family of exponential regulators, and for definiteness we only discuss projection scheme A.
As an example, the dependence of $\theta_1$ and $\eta$ on $a$ is plotted in \autoref{fig:HB_NLO_opt}. It seems that the regulator dependence
is a bit stronger compared to the one in the Ising model \cite{Knorr:2016sfs}. Still, optimised values for the first two critical exponents
and the anomalous dimension can be inferred with \PMS{},
\begin{align}
 \theta_1^\text{opt} &= 1.4178 \, , \qquad &a^\text{opt} = 1.66 \, , \notag \\
 \theta_2^\text{opt} &= -0.7473 \, , \qquad &a^\text{opt} = 1.73 \, , \notag \\
 \eta^\text{opt} &= 0.04662 \, , \qquad &a^\text{opt} = 1.57 \, .
\end{align}
The optimal values for the parameter $a$ are quite close to the ones in the Ising model \cite{Knorr:2016sfs}, and consistently optimise the first two critical 
exponents as well as the anomalous dimension. More general optimisation criteria can be found in Refs. \onlinecite{Litim:2000ci, Litim:2001up, Litim:2001fd, 
Pawlowski:2005xe, Pawlowski:2015mlf}. For comparison, we give recent Monte Carlo\cite{PhysRevB.84.125136} (\MC{}) and conformal bootstrap\cite{Kos2016} 
(\CBS{}) results:
\begin{equation}
\begin{aligned}
 \theta_1^\text{MC} &= 1.4053(20) \, , \\
 \eta^\text{MC} &= 0.0378(3) \, , \\
 \theta_1^\text{CBS} &= 1.4043(55) \, , \\
 \eta^\text{CBS} &= 0.03856(124) \, .
\end{aligned}
\end{equation}
The optimised value for the leading critical exponent is in good agreement with these estimates, differing by only 1\%. As expected, the anomalous dimension 
needs further improvement by enhancing the truncation. In comparison to a truncation that only retains the potential and the anomalous dimension 
\cite{Classen:2015mar}, $\theta_1$ changes by about $4\%$.

\section{Results for the chiral Heisenberg model}\label{sec:GN}

Let us now switch on fermions, and study the full system \eqref{eq:HGNansatz} at criticality.
The fixed point lies in the symmetric regime, and thus, schemes A and B are the same. For definiteness,
we only discuss the family of exponential regulators.

\begin{figure*}
 \includegraphics[width=0.944\textwidth]{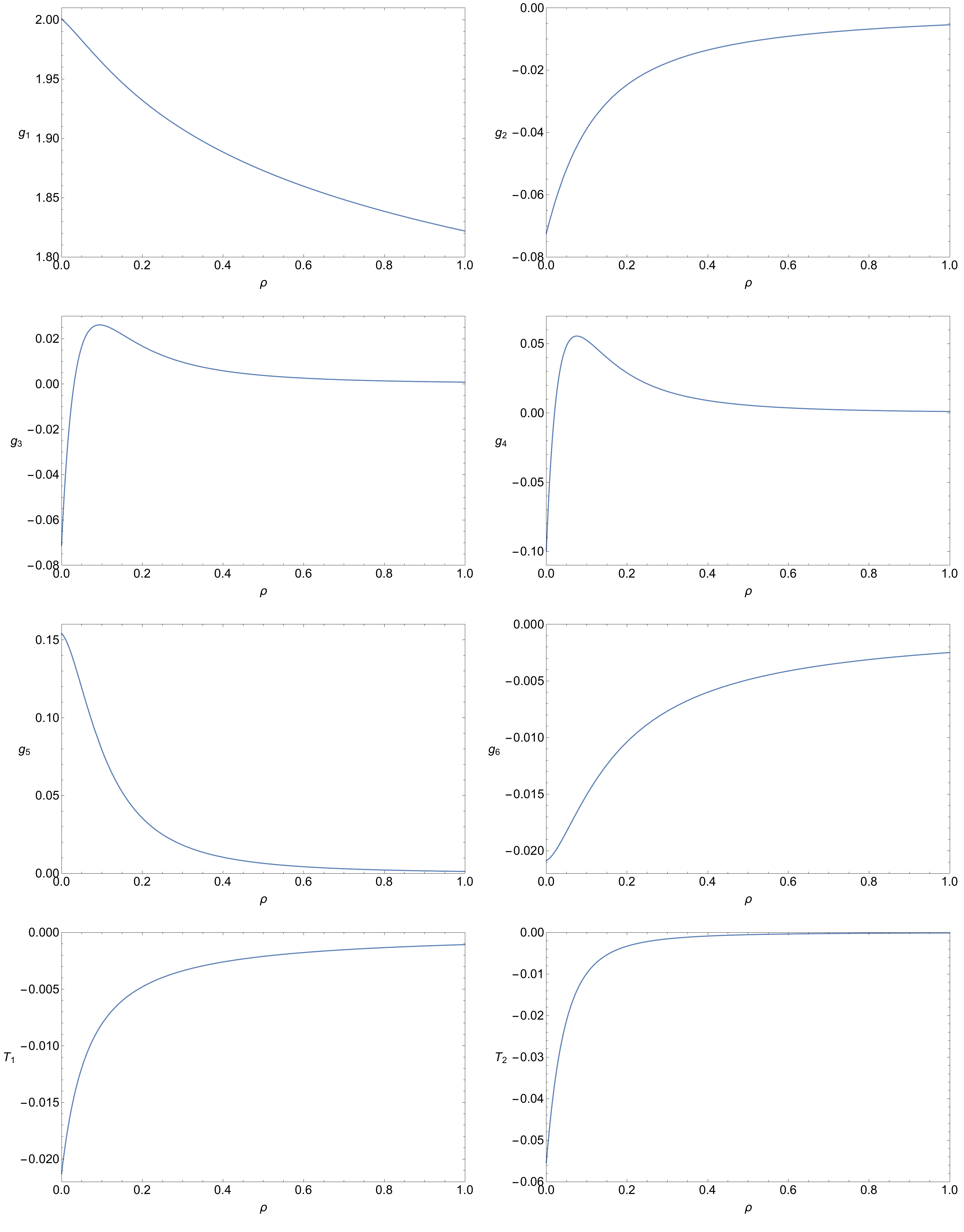}
 \caption{Fixed point solution of the generalised Yukawa, and tensorial interactions of the chiral Heisenberg model, for the
 regulator parameter $a=2$.
 Since $\eta_\phi$ is larger than $1$, the standard Yukawa coupling
 $g_1$ falls off to zero for large $\rho$, in contrast to the Yukawa coupling of the chiral Ising model.
 The other couplings are suppressed, as expected from their high mass dimension.}
 \label{fig:cHB2}
\end{figure*}

As exemplary case, we show the fixed point solution for the specific regulator parameter choice $a=2$ in \autoref{fig:cHB1} and
\autoref{fig:cHB2}. The first of the two figures displays the first derivative of the potential, the two wave function renormalisations
and the correction term to the radial propagator of the bosons. Since $V'>0$, we are in the symmetric regime. In contrast to the
purely bosonic case, $Y_\phi$ is negative. This is not a problem, since the combination of $Z_\phi$ and $Y_\phi$ that appears in the propagator
is strictly positive. \autoref{fig:cHB2} shows the standard as well as the generalised Yukawa interactions. As the bosonic anomalous dimension
is larger than $1$, the Yukawa interaction decreases for increasing $\rho$, contrary to the case of the standard Gross-Neveu model \cite{Knorr:2016sfs}.
All generalised Yukawa couplings and the tensorial couplings are suppressed, as their mass dimension suggests.

\begin{figure*}
 \includegraphics[width=\textwidth]{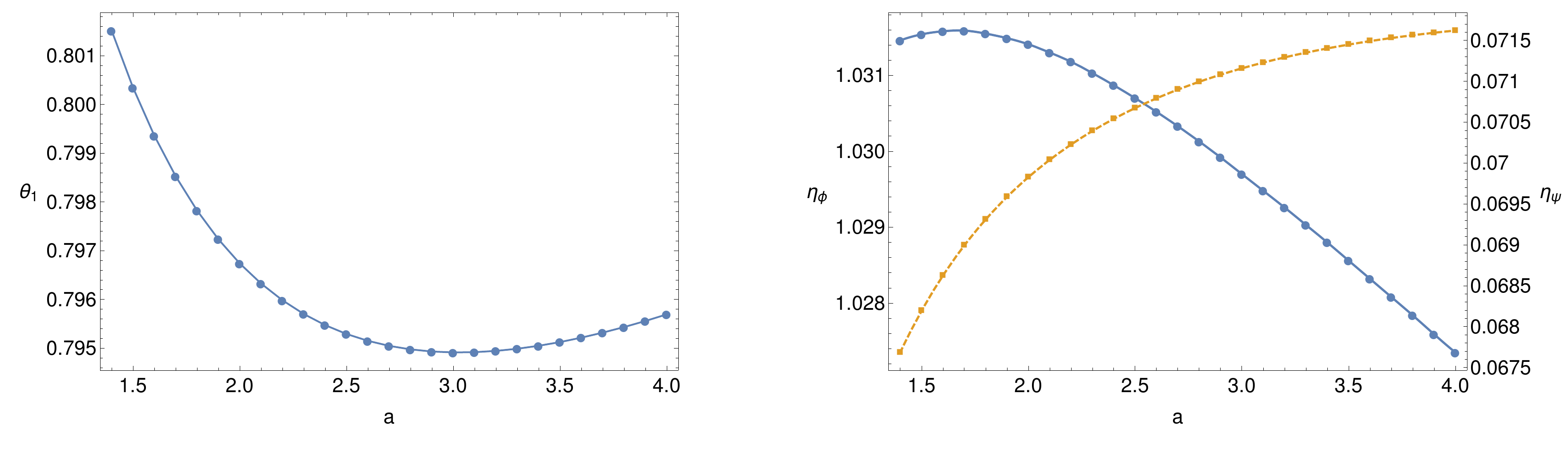}
 \caption{Regulator dependence of physical quantities of the chiral Heisenberg model.
 On the left panel, the first critical exponent is shown. On the right panel,
 the bosonic (blue dots) and fermionic (orange boxes) anomalous dimensions are plotted.
 Interpolations help to guide the eye. Whereas both the critical exponent and the bosonic
 anomalous dimension show an extremum, the fermionic anomalous dimension doesn't, and cannot be
 optimised by \PMS{}.}
 \label{fig:GN_NLO_regscan}
\end{figure*}

From the dependence on the regulator parameter $a$, we can optimise our estimates for physical quantities. The dependence
of the first critical exponent and the two anomalous dimensions on $a$ is shown in \autoref{fig:GN_NLO_regscan}. From that, we obtain the
optimised values
\begin{align}
 \theta^\text{opt} &= 0.795 \, , \qquad &a^\text{opt} = 3.03 \, , \notag \\
 \eta_\phi^\text{opt} &= 1.032 \, , \qquad &a^\text{opt} = 1.68 \, .
\end{align}
The fermionic anomalous dimension doesn't display an extremum in the considered parameter range. Taking the value at $a=3$ and accounting for the dependence on 
$a$ by varying it by $\pm 1$, we estimate
\begin{equation}
 \eta_\psi = 0.071(2) \, .
\end{equation}
\autoref{tab:litGN} shows a comparison to results obtained by different methods. The present results are in good agreement
with earlier \FRG{} studies that only retain the field dependence of the potential together with field-independent wave functions
and standard Yukawa coupling $g_1$ \cite{Janssen:2014gea}. The critical exponent and the bosonic anomalous dimension only change
by a few percent, the fermionic anomalous dimension by about 17$\%$. Our results are in some tension with results obtained with an
$\epsilon$-expansion \cite{Rosenstein:1993zf, Janssen:2014gea, Zerf:2017zqi}, and in even stronger disagreement with quantum Monte Carlo 
results \cite{Toldin:2014sxa, Otsuka:2015iba}. Taking our results as point of reference, the Monte Carlo estimates for the critical exponent disagree by $23\%$ and 
$50\%$, respectively. For the bosonic anomalous dimension, the discrepancy is about $32\%$ and $53\%$, respectively, whereas the fermionic anomalous dimension 
from the quantum Monte Carlo simulation of Ref. \onlinecite{Otsuka:2015iba} differs by about a factor of $3$. This raises the question whether the same 
universality class is studied.
\begin{table}
\begin{tabular}{cccc}
\hline\hline
  & $\theta_1$ & $\eta_\phi$ & $\eta_\psi$ \\ \hline
  FRG (this work) & 0.795 & 1.032 & 0.071(2) \\
  FRG \cite{Janssen:2014gea} & 0.772 & 1.015 & 0.084 \\
  quantum Monte Carlo \cite{Toldin:2014sxa} & 1.19(6) & 0.70(15) & \textemdash \\
  quantum Monte Carlo \cite{Otsuka:2015iba} & 0.98(1) & 0.49(5) & 0.20(2) \\
  $(4-\epsilon)$ 2nd order \cite{Rosenstein:1993zf, Janssen:2014gea} & 0.834 & 0.959 & 0.242 \\
  $(4-\epsilon)$ 4th order (3/1) Pad\'e \cite{Zerf:2017zqi} & 0.645 & 0.956 & 0.156 \\ \hline\hline
\end{tabular}
\caption{Comparison of the first critical exponent and the anomalous dimensions of the chiral Heisenberg model with the literature. To estimate the bosonic 
anomalous dimension from the results of Ref. \onlinecite{Otsuka:2015iba}, we employed the hyperscaling relation $\eta_\phi = 2\beta\theta_1+2-d$.}
\label{tab:litGN}
\end{table}

\section{Summary}\label{sec:summary}

The present work completes our study of the two subsystems of the particular model of graphene put forward in Refs. \onlinecite{Roy:2011pg, Roy:2013aya, 
Classen:2015mar}, including the presumably most important operators at next-to-leading order in the derivative expansion. In the first part, we investigated 
the model without fermions, \ie{} the Heisenberg model. The ambiguity in the normalisation of the wave function renormalisation was shown to be not a
problem at all, rather both schemes that have been investigated deliver a consistent picture of the model at criticality. This is in contrast to the Ising model
at the same level of truncation, and might indicate that $O(N)$-symmetric models with $N=3$ are already well described by a large-$N$ approximation,
as there the influence of the wave function renormalisation is parametrically suppressed.

Regarding the chiral model, we studied a very extensive truncation, including 12 operators. In particular, we focussed on the momentum-dependent corrections 
to the standard Yukawa coupling, and discussed the effect of tensorial couplings. The stability of critical quantities, as the leading critical exponent or the 
anomalous dimensions, is remarkable, if one compares to a truncation that only retains minimal information, similarly to the case of the chiral Ising model 
\cite{Knorr:2016sfs}. This is taken as a strong hint to an exceptionally good convergence behaviour of the derivative expansion in such models, and strengthens 
our trust in the quantitative accuracy of the present results.

The previously found disagreement \cite{Janssen:2014gea} with quantum Monte Carlo studies \cite{Toldin:2014sxa} persists, and now includes also the more recent 
quantum Monte Carlo results of Ref. \onlinecite{Otsuka:2015iba}. The first critical exponent and the bosonic anomalous dimension differ by $20$ to $50\%$, the 
fermionic anomalous dimension disagrees by a factor of about $3$. The present work suggests that including more operators with more derivatives won't change the 
results by a lot. By comparison with results from both the $\epsilon$-expansion \cite{Rosenstein:1993zf, Janssen:2014gea, Zerf:2017zqi} and quantum Monte Carlo studies 
\cite{Otsuka:2015iba}, we however expect the fermionic anomalous dimension to become larger upon inclusion of further operators. Let us note that if the Monte Carlo
investigations are indeed in a different universality class, then the investigations with the \FRG{} present to date the only nonperturbative results on the chiral Heisenberg
model.

There are at least two possibilities for improvement of the present calculation. On the one hand, four-fermion terms can be included, ideally by a 
dynamical bosonisation along the lines of Refs. \onlinecite{Gies:2001nw, Jaeckel:2002rm, Pawlowski:2005xe, Gies:2006wv, Floerchinger:2009uf, 
Floerchinger:2009pg, Janssen:2012pq}. On the other hand, to go beyond the derivative expansion, also the momentum dependence can be resolved, see \eg{} Refs. 
\onlinecite{Benitez:2009xg, Benitez:2011xx} for works resolving both momentum and field dependences with the \FRG{}.

In principle, now we are in the situation to study the combined system of \eqref{eq:HGNansatz} and Ref. \onlinecite{Knorr:2016sfs}, and
give precision estimates on critical quantities for Dirac materials. Unfortunately, even for the uncoupled fixed points that can be constructed
directly from the solutions of the two subsystems, no estimate on the decisive third critical exponent can be made, as no scaling relation is known,
in contrast to the situation in \eg{} the $O(N)\oplus O(M)$-model. Also, further operators appear that mix the two bosonic order parameters, and these will 
be important both for the determination of the third critical exponent of the uncoupled fixed points, and for the determination of fully coupled fixed points. 
Still, both the technology put forward and the experience with the subsystems will be helpful to study 
the coupled system. In particular, it seems that a truncation which resolves the potential, all kinetic terms of the bosons, the kinetic term of the fermions 
and the standard Yukawa coupling are already reliable quantitatively.

\section*{Acknowledgements}

I would like to thank J. Borchardt, L. Classen, H. Gies, B. Ihrig, S. Lippoldt, M. M. Scherer and A. Wipf
for useful discussions during different stages of this project and H. Gies
for valuable comments on the manuscript. This work was supported by the Deutsche
Forschungsgemeinschaft (DFG) graduate school ``Quantum and Gravitational Fields'' GRK 1523/2,
and by the DFG grant no. Wi 777/11.

\bibliography{specbib}

\end{document}